\documentclass[twocolumn]{aastex62}

\graphicspath{{./}{figures/}}

\received{xxx}
\revised{xxx}
\accepted{xxx}
\submitjournal{ApJ}

\shorttitle{Fluid Love number $k_2$ in exoplanetary transits}
\shortauthors{Hellard et al.}

\begin{document}

\title{Retrieval of the fluid Love number $k_2$ in exoplanetary transit curves}

\correspondingauthor{Hugo Hellard}
\email{hugo.hellard@dlr.de}

\author[0000-0003-0077-3196]{Hugo Hellard}
\affil{Deutsches Zentrum f\"ur Luft- und Raumfahrt \\
Rutherfordstra\ss e 2, 12489
Berlin, Germany}
\affiliation{Technische Universit\"at Berlin \\
Stra\ss e des 17. Juni 135, 10623
Berlin, Germany}

\author{Szil\'{a}rd Csizmadia}
\affiliation{Deutsches Zentrum f\"ur Luft- und Raumfahrt \\
Rutherfordstra\ss e 2, 12489
Berlin, Germany}

\author{Sebastiano Padovan}
\affiliation{Deutsches Zentrum f\"ur Luft- und Raumfahrt \\
Rutherfordstra\ss e 2, 12489
Berlin, Germany}

\author{Heike Rauer}
\affiliation{Deutsches Zentrum f\"ur Luft- und Raumfahrt \\
Rutherfordstra\ss e 2, 12489
Berlin, Germany}
\affiliation{Technische Universit\"at Berlin \\
Stra\ss e des 17. Juni 135, 10623
Berlin, Germany}
\affiliation{Freie Universit\"at Berlin \\
Kaiserswertherstra\ss e 16-18, 14195
Berlin, Germany}

\author{Juan Cabrera}
\affiliation{Deutsches Zentrum f\"ur Luft- und Raumfahrt \\
Rutherfordstra\ss e 2, 12489
Berlin, Germany}

\author{Frank Sohl}
\affiliation{Deutsches Zentrum f\"ur Luft- und Raumfahrt \\
Rutherfordstra\ss e 2, 12489
Berlin, Germany}

\author{Tilman Spohn}
\affiliation{Deutsches Zentrum f\"ur Luft- und Raumfahrt \\
Rutherfordstra\ss e 2, 12489
Berlin, Germany}

\author{Doris Breuer}
\affiliation{Deutsches Zentrum f\"ur Luft- und Raumfahrt \\
Rutherfordstra\ss e 2, 12489
Berlin, Germany}



\begin{abstract}
We are witness to a great and increasing interest in internal structure, composition and evolution of exoplanets. However, direct measurements of exoplanetary mass and radius cannot be uniquely interpreted in terms of interior structure, justifying the need for an additional observable. The second degree fluid Love number, $k_2$, is proportional to the concentration of mass towards the body's center, hence providing valuable additional information about the internal structure. When hydrostatic equilibrium is assumed for the planetary interior, $k_2$ is a direct function of the planetary shape. Previous attempts were made to link the observed tidally and rotationally induced planetary oblateness in photometric light curves to $k_2$ using ellipsoidal shape models. Here, we construct an analytical 3D shape model function of the \textit{true} planetary mean radius, that properly accounts for tidal and rotational deformations. Measuring the \textit{true} planetary mean radius is critical when one wishes to compare the measured $k_2$ to interior theoretical expectations. We illustrate the feasibility of our method and show, by applying a Differential Evolution Markov Chain to synthetic data of WASP-121b, that a precision $\leq$ 65 ppm/$\sqrt{2\,min}$ is required to reliably retrieve $k_2$ with present understanding of stellar limb darkening, therefore improving recent results based on ellipsoidal shape models. Any improvement on stellar limb darkening would increase such performance.
\end{abstract}



\section{Introduction} \label{sec:intro}

Knowledge of the planetary radius and mass is not sufficient to infer the interior structure, as different composition and density profiles can lead to the same solution (\textit{e.g.} \citealt{rogers2010}), showing the need for additional observables. About 50\% of the about 2800 confirmed transiting exoplanets orbit their host star in less than 10 days \footnote{\url{http://exoplanet.eu}, \citet{schneider2011}}. Such close-in objects undergo strong tidal and rotational deformations, which modify their shapes from spherical to more complicated ones. We try to interpret the shape in terms of the internal structure parameters $k_j$, where $k_j$ is the fluid Love number of degree $j$, introduced by \citet{love1911}. Therefore, we present our recently developed model in which we assume a spherical host star and a planet in hydrostatic equilibrium, \textit{i.e.}, behaving as a fluid at rest. In particular, the second-degree Love number, $k_2$, is an indication of mass concentration towards the body's center, providing additional information about the interior (see, \textit{e.g.}, \citealt{kellermann2018}). A value of $k_2 = 0$ indicates a mass-point approximation (a.k.a., Roche model), $k_2 = 1.5$ corresponds to a fully homogeneous body, and its full derivation depends on the internal radial density profile, (see, \textit{e.g.}, \citealt{padovan2018}). As a result of tidal and rotational deformations, the stellar eclipsed area during transit will differ from a transiting sphere, modifying the corresponding transit light curve \citep{hui2002, seager2002}. Hence, one should be able to improve our understanding of planetary internal structure by measuring the planetary shape from transit light curves.\\
Previous work focused on transit light curves of oblate planets, defined either through two-axes ellipsoids (\textit{e.g.} \citealt{carter2010}) or three-axes ellipsoids (\textit{e.g.} \citealt{correia2014,maxted2016,akinsanmi2018}). When compared to a transiting sphere, distortions in photometric light curves were identified mainly at ingress and egress phases of the transit due to polar oblateness. The latter, often called flattening, was then related to the interior through $k_2$ or through the quadrupole moment $J_2$ \citep{carter2010,correia2014}).\\
In this paper, we first construct a three-dimensional analytical shape model based on the theories of \citet{love1911} and \citet{kopal1959}, that properly accounts for tidal and rotational deformations. This model presents the advantage of depending on the \textit{true} planetary mean radius, an essential parameter when computing theoretical $k_2$ based on interior models (see, \textit{e.g.}, \citealt{padovan2018}). The \textit{true} planetary mean radius is the radius $R_p$ as defined in Equation (\ref{eq:3dshape}). Thus, it corresponds to the (spherical) radius the planet would have if it were isolated and non-rotating. We also show that ultra-short period planets or planets with potentially fast rotation rates are the most suitable targets to constrain $k_2$ from photometric transit light curves. We then describe our algorithm to compute transit light curves and apply a Differential Evolution Markov Chain \citep{terbraak2008} to illustrate the feasibility of measuring exoplanetary shapes by taking the example of WASP-121b in the light of dedicated space missions.

\section{Shape model}
We assume circular orbits and do not account for interactions between rotation and tides, which are in general terms of second order or higher \citep{landin2009}. Below we derive the planetary surface shape (subscript $p$), and one can derive the stellar shape (subscript $s$) by an obvious interchange of signs. We define the usual Cartesian coordinate system, with its origin at the planet's center of mass, the x-axis pointing towards the star's center of mass, the z-axis perpendicular to the orbital plane, and the y-axis to form a right-handed system. We define a spherical coordinate system, with the same origin, where at point $P$ the coordinates are radius $r$, colatitude $\theta$, and east longitude $\phi$, so that
\begin{eqnarray}
x &=& r\lambda = r\sin\theta\cos\phi, \\
y &=& r\mu = r\sin\theta\sin\phi, \\
z &=& r\nu = r\cos\theta.
\end{eqnarray}
The tidal potential at the planetary surface, $V_t$, can be expanded in spherical harmonics as (\textit{e.g.} \citealt{kopal1959})
\begin{equation}
V_t = \frac{Gm_{s}}{d}\sum_{j=2}^{\infty}\left(\frac{R_p}{d}\right)^{j}P_{j}(\lambda),
\label{eq:tide_potential}
\end{equation}
where $G$ is the gravitational constant, $m_s$ is the stellar mass, $d$ is the orbital semi-major axis, $R_p$ is the \textit{true} planetary mean radius, and $P_j$ is the Legendre polynomial of degree $j$. \citet{kopal1959} showed that omitting terms with degree $j \leq 4$ is equivalent to considering the Roche model, which describes the behavior of a mass-point surrounded by a massless envelope (see, \textit{e.g.}, \citealt{wilson1971}).\\
We consider the general case where the spin axis is not perpendicular to the orbital plane. We define the obliquity, $\Theta$, as the angle between the arbitrary radius vector given by $\lambda, \mu, \nu$ and the spin axis whose direction is given by
\begin{eqnarray}
\lambda^{'} &=& \sin\beta\cos\alpha, \\
\mu^{'} &=& \sin\beta\sin\alpha, \\
\nu^{'} &=& \cos\beta.
\end{eqnarray}
where $\alpha$ and $\beta$ are the longitude and co-latitude of the spin axis, respectively. Hence, $\Theta$ is given by
\begin{equation}
\cos(\Theta) = \lambda\lambda^{'} + \mu\mu^{'} + \nu\nu^{'}.
\end{equation}
\citet{kopal1959} showed that the rotational potential, $V_r$, at the planet's surface can be written as
\begin{equation}
V_r = -\frac{1}{3}\frac{Gm_p}{d}(1+q)F_p^2\left(\frac{R_p}{d}\right)^2P_{2}(\cos(\Theta)),
\label{eq:rot_potential}
\end{equation}
where $q = m_s/m_p$ is the mass ratio and $F_{p} = P_{orb}/P_{rot}$ is the ratio between the orbital and rotational periods.\\
\\
\citet{love1911} demonstrated that the radial surface deformation of degree j, $s_{j}$, is given by
\begin{equation}
R_ps_j = \frac{1+k_j}{g}V_{j}(r=R_p),
\label{eq:SurfLove}
\end{equation}
where $k_j$ is the potential fluid Love number of degree $j$, g is the gravitational surface acceleration, and $V_j$ is a perturbing potential of degree $j$ calculated at the planet's mean radius $R_p$. In the fluid approximation $h_j = 1+k_j$, where $h_j$ is the surface radial deformation fluid Love number. According to Equations (\ref{eq:tide_potential}) and (\ref{eq:SurfLove}) we obtain the tidal surface deformations
\begin{equation}
s_t = q\sum_{j=2}^{4}h_jP_j(\lambda)\left(\frac{R_p}{d}\right)^{j+1}.
\label{eq:tides}
\end{equation}
The rotational surface deformations are also found according to Equations (\ref{eq:rot_potential}) and (\ref{eq:SurfLove})
\begin{equation}
s_r = -\frac{1}{3}h_2(1+q)F_p^2\left(\frac{R_p}{d}\right)^3P_{2}(\cos(\Theta)),
\label{eq:rot}
\end{equation}
Assuming that the surface deformations are simply additive, the total surface shape is defined by
\begin{equation}
r(\theta,\phi) = R_p\left(1 + \sum_{j}s_{j}\right),
\end{equation}
leading to (with Equations (\ref{eq:tides}) and (\ref{eq:rot}))
\begin{eqnarray}
\label{eq:3dshape}
r(\theta,\phi) &=& R_p\Bigg(1 + q\sum_{j=2}^{4}h_jP_j(\lambda)\left(\frac{R_p}{d}\right)^{j+1} \\
\nonumber
&-& \frac{1}{3}h_2(1+q)F_p^2\left(\frac{R_p}{d}\right)^3P_{2}(\cos(\Theta))\Bigg).
\end{eqnarray}
We note here that the fluid Love number $k_j$ ($h_j = 1+k_j$) is twice the apsidal motion constant in \citet{kopal1959} (also indicated as $k_j$) and that $h_j$ is equal to the parameter $\Delta_j$ in \citet{kopal1959}. The effect of the interior is entirely enclosed in the fluid Love numbers $k_j$, or equivalently $h_j$.\\
Equation (\ref{eq:3dshape}) also applies to the star by switching the origin of the coordinate system at the stellar center of mass, by interchanging the subscript $p$ with $s$ (including in the mass ratio $q$), and by taking the stellar fluid Love numbers (see, \textit{e.g.}, \citealt{claret2004}).\\
We validate our model by verifying that for $k_{2,3,4} = 0$, we obtain the Roche model. Additionally, we retrieve the shape of solar system planets that exhibit a polar flattening (\textit{e.g.} the Earth and Jupiter for which we considered $k_2 = 0.934$ \citet{lambeck1980} and $k_2 = 0.537$ \citet{wahl2016}, respectively).\\
\\
Our three-dimensional model properly accounts for tidal and rotational deformations and does not require any approximations except hydrostatic equilibrium in the interior and absence of non-linear effects in the planetary response to perturbations. It allows to directly fit the \textit{true} planetary mean radius, the fluid Love numbers, the rotational period (enclosed in $F_p$), and the spin obliquity from observations. This is a major improvement compared to previous and recent works (\textit{e.g.} \citealt{carter2010, correia2014, zhu2014, maxted2016, akinsanmi2018}). The derivation of the \textit{true} planetary mean radius is critical as it enters the calculation of theoretically expected $k_2$.\\
\\
The shape modeling adopted in this paper differs from ellipsoidal shape models, leading to different planetary deformations. These differences appear in turn in transit light curves and, as a result, the noise level at which one can safely measure $k_2$ will differ, as reported in Section \ref{sec:discussion}. In Appendix \ref{sec:shapemodels} we compare some planetary shapes calculated by \citet{correia2014} with those obtained using our model. \\
\\
The larger the planetary surface deformations, the easier the retrieval of $k_2$ from transit light curves. The dimensionless absolute quantities that give gravitational acceleration in comparison to that of rotational and tidal are respectively given by \citep{kellermann2018}
\begin{equation}
q_r = F_p^2(1+q)\left(\frac{R_p}{d}\right)^3 \quad , \quad q_t = 3q\left(\frac{R_p}{d}\right)^3.
\end{equation}
Hence, surface deformations for synchronously locked close-in planets are tidally dominated ($q_t / q_r \simeq 3$), while for outer planets with fast spin rates, surface deformations are rotationally dominated ($q_r / q_t \simeq F_p^2 / 3$), for instance Jupiter and Saturn or see \citet{zhu2014} for exoplanets. Therefore, planets orbiting close to their Roche limit - the distance at which the tidal forces will overcome the planet's gravity and lead to its tidal disruption - (synchronously locked or not) and outer planets with fast rotation rates are the best candidates to measure their fluid Love number $k_2$.

\section{Retrieval of $\MakeLowercase{k}_2$ in transit light curves}
\label{sec:retrieval_method}

For the sake of illustration and based on the criteria mentioned previously, we consider the object WASP-121b, a hot Jupiter with a mean density of $\rho = 0.201 \pm 0.010 \, \rho_{J}$ orbiting at roughly twice its Roche limit. We assume a planetary synchronous rotation and a spherical star with a quadratic limb darkening law ($u_1 = u_2 = 0.3$) \citep{claret2011}. Our model, however, allows the user to choose between several limb darkening laws so that the preferred one (based on, \textit{e.g.}, chi-square or Bayesian information) can be adopted. The main assumed properties of the planet and its host star are summarized in Table \ref{tab:planets}. We emphasize that the assumed radius ratio $R_p/R_s$ given in Table \ref{tab:planets} corresponds to the \textit{true} normalized planetary mean radius in our shape model, \textit{a priori} unknown. This radius differs from the one retrieved through a spherical fitting procedure. We considered a planetary Love number $k_2 = 0.5$ based on recent models \citep{wahl2016}, and forced $k_3 = k_4 = 0$. Indeed, \citet{padovan2018} showed that $k_2$ has the strongest dependence on central mass concentration among the three. We highlight that only synthetic light curves were created to prove the feasibility of our method, not real data, thus our results should not be physically interpreted in terms of, \textit{e.g.}, internal structure.

\begin{deluxetable}{c|c}[ht!]
\tablecaption{Main assumed properties of the WASP-121 system \citep{claret2011,delrez2016}. The value of $k_2$ is arbitrarily chosen. \label{tab:planets}}
\tablehead{
\colhead{Parameter (unit)} & \colhead{Assumed value}
}
\startdata
$m_{s}$ ($M_{\odot}$) & 1.353 \\
$Vmag$ & 10.4 \\
$u_1$ & 0.3 \\
$u_2$ & 0.3 \\
$m_{p}$ ($M_{J}$) & 1.184 \\
Inclination (deg) & 87.6 \\
Eccentricity & 0.0 \\
$\frac{R_p}{R_s}$ & 0.1313 \\
$\frac{d}{R_s}$ & 3.7486 \\
$k_2$ & 0.5
\enddata
\end{deluxetable}

\subsection{Forward modeling} \label{subsec:forward}

\begin{figure}
\plotone{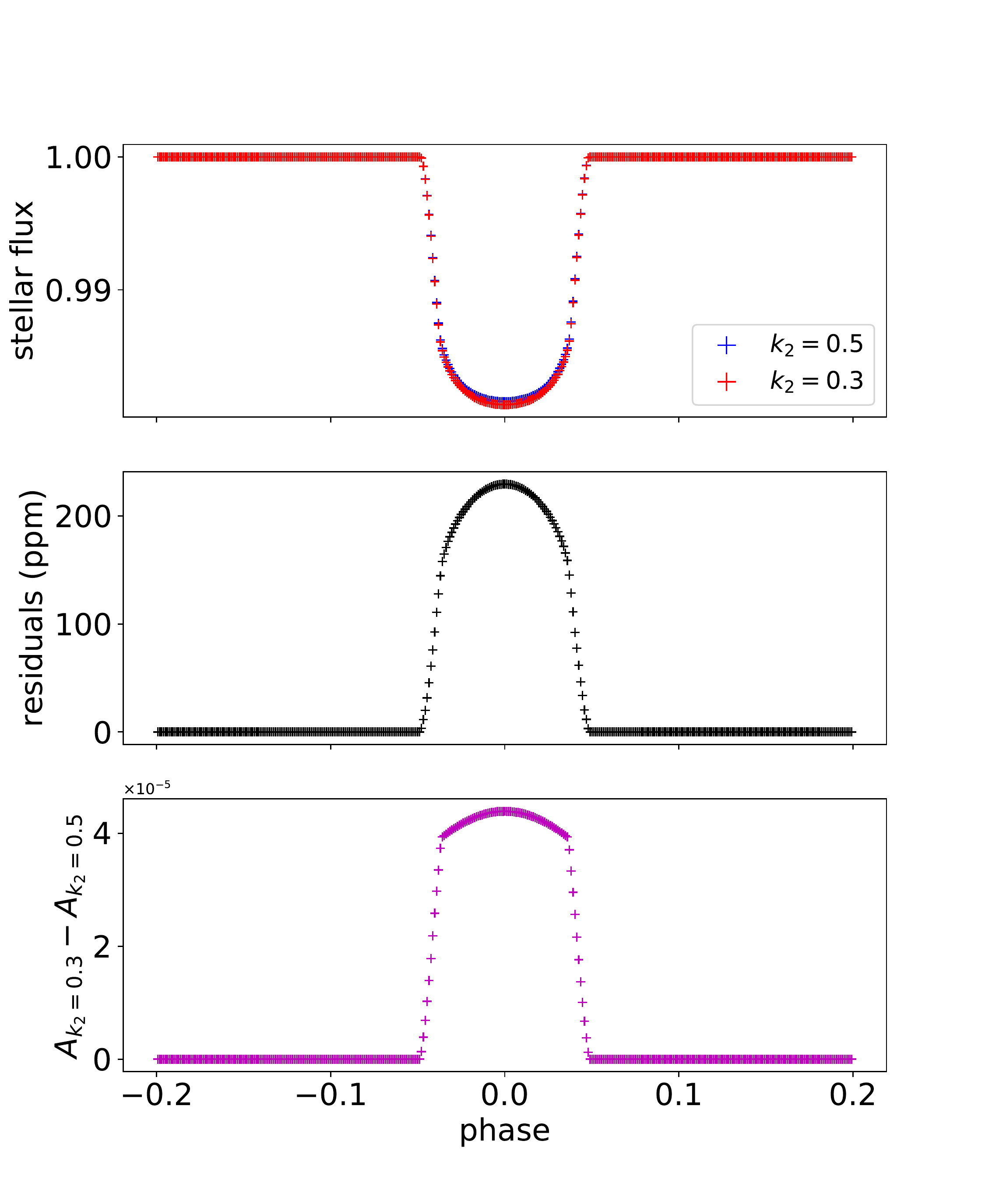}
\caption{WASP-121b modeled light curves for $k_2=0.3$ and $k_2=0.5$ (top), their difference (middle), and the difference in the normalized stellar eclipsed area (bottom).}
\label{fig:light_curves_diff}
\end{figure}

To derive a single transit event, the stellar eclipsed area is derived by projecting the planetary shape onto the plane-of-sky. We define the plane-of-sky coordinate system, centered at the star's center of mass, with the x-axis pointing towards the observer, the z-axis perpendicular to the x-axis and pointing up, and the y-axis to form a right-handed system. By denoting $i$ the orbital inclination, and $\varphi$ the orbital phase, the y and z sky-projected planetary coordinates normalized to the semi-major axis, $y_{sky}$ and $z_{sky}$, are given by
\begin{eqnarray}
y_{sky}(\varphi) &=& x\sin\varphi - y\cos\varphi + \sin\varphi \\
z_{sky}(\varphi) &=& x\cos i \cos\varphi + y\cos i \sin\varphi + z\sin i \\
\nonumber
		&-& \cos i \cos\varphi
\end{eqnarray}
The stellar eclipsed area is simply the overlapping contour of the sky-projected planetary shape with the stellar spherical contour. This area being non-analytical, we uniformly sample it using a two dimensional Sobol sequence \citep{sobol1999} and the stellar flux deficit is computed using a Monte-Carlo integration. The latter takes most of the computing time as a high precision on the eclipsed area is required. After implementing our model in Python, one transit simulation took about 450 ms with a precision of 1 ppm on a 2.40 GHz Intel Core i7 Toshiba Satellite (running on Ubuntu 18.04 LTS).\\
\\
For illustration purpose, we present in Figure \ref{fig:light_curves_diff} two modeled transit light curves for $k_2=0.3$ and $k_2=0.5$, as well as their difference, and the difference in the stellar eclipsed area. It is clear from Figure \ref{fig:light_curves_diff} (top panel) that the planetary mean radius and $k_2$ are degenerate as a larger radius could explain most of the difference between both light curves. It is essential to note that the whole transit has to be carefully monitored to precisely retrieve $k_2$. Therefore, the knowledge of the stellar limb darkening will be critical (see Section \ref{sec:discussion}).\\

\begin{figure}[ht!]
\plotone{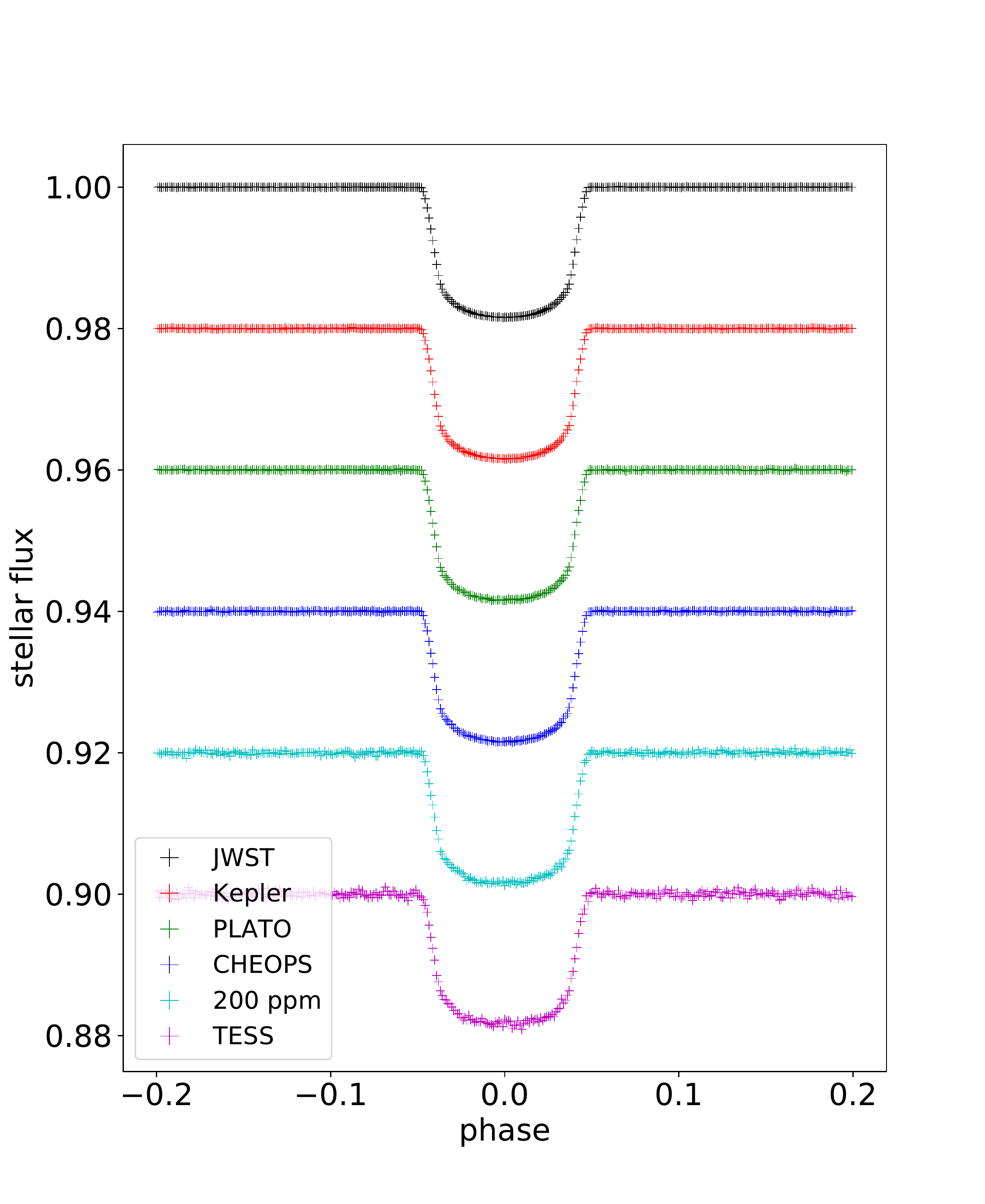}
\caption{Synthetic transit light curves of WASP-121b with $k_2 = 0.5$ for different observing facilities, with white noise as described in Table \ref{tab:noise}. The light curves were shifted for clarity.}
\label{fig:lcs}
\end{figure}

To be able to measure the planetary shape in transit light curves, we need a good enough time resolution and low noise (as shown in Figure \ref{fig:light_curves_diff}). Therefore, we calculate photometric white noise levels (\textit{i.e.} photon noise), $\sigma$, for a composite light curve from 10 observed transits, binned into 2-min intervals, for several observing facilities. For a given noise level, $\sigma_{ref}$, corresponding to a reference time $t_{ref}$, we calculate $\sigma$ using
\begin{equation}
\sigma = \sigma_{ref}\,\sqrt[]{\frac{t_{ref}}{2\,min}}\,\sqrt[]{\frac{1}{10}},
\label{eq:noise}
\end{equation}
since photon noise is equal to the square root of the number of photons received during $t_{ref}$. The values are reported in Table \ref{tab:noise}. We considered an additional noise level of 200 ppm/$\sqrt{2min}$ to have a value between those of CHEOPS and TESS. These white noise levels are then randomly added using a simple Gaussian distribution to create synthetic light curves for each observing facility. We create three realizations of noisy light curve for each facility, and report the average values of these realizations. An example is presented in Figure \ref{fig:lcs}.

\begin{deluxetable*}{c|cc|c}
\tablecaption{Photometric white noise levels ($\sigma$) for a composite transit light curve of WASP-121b from 10 observed transits, binned into 2-min intervals, for several observing facilities (equation \ref{eq:noise}). \label{tab:noise}}
\tablehead{
\colhead{Facility} & \colhead{$t_{ref}$ (min)} & \colhead{$\sigma_{ref}$ (ppm/$\,\sqrt[]{t_{ref}}$)} &
\colhead{$\sigma$ (ppm/$\,\sqrt[]{2\,min}$) for 10 transits}
}
\startdata
TESS \tablenotemark{a} & 60 & 208 & 360\\
CHEOPS \tablenotemark{b} & 1 & 319 & 71\\
PLATO \tablenotemark{c} & 60 & 34 & 63 \\
Kepler \tablenotemark{d} & 1 & 202 & 45\\
JWST (NIRSpec) \tablenotemark{e} & \textit{1 sec} & 780 & 23
\enddata
\tablecomments{The initial reference noise levels ($\sigma_{ref}$) correspond to a specified reference time ($t_{ref}$) @ 10.4 $Vmag$ for each facility.}
\tablenotetext{a}{\url{https://heasarc.gsfc.nasa.gov/cgi-bin/tess/webtess/wtv.py}}
\tablenotetext{b}{\citet{akinsanmi2018}}
\tablenotetext{c}{PLATO Definition Study Report, ESA-SCI(2017)1, April 2017. We applied an additional 10\% margin.}
\tablenotetext{d}{\url{https://keplerscience.arc.nasa.gov/}}
\tablenotetext{e}{\url{https://jwst.etc.stsci.edu/}}
\end{deluxetable*}

\subsection{Retrieval of $k_2$} \label{subsec:retrieval}

The parametrization for transit fitting has been done in different ways in the literature. We decide to fit the inclination, $i$, the sum and difference of the quadratic limb darkening coefficients $u_p = u_1 + u_2$ and $u_m = u_1 - u_2$, the normalized semi-major axis (in stellar radii) $d/R_s$, the normalized \textit{true} mean planetary radius (in stellar radii) $R_p/R_s$, and the second degree fluid Love number $k_2$. We forced the third and fourth degree planetary fluid Love numbers to 0. By assuming the probability distribution of uncertainties to be Gaussian, and assuming uncorrelated uncertainties, the quantity to be minimized is $\chi^2$ and given by
\begin{equation}
\chi^{2} = \sum_{i=1}^{n}\left(\frac{O_i-M_i}{\sigma}\right)^2,
\end{equation}
where $O$ and $M$ are the observed and modeled data sets, of $n$ data points, respectively. We robustly find a local minimum employing the downhill simplex method \citep{nelder1965}. We then wish to estimate the goodness of the fit by performing a Markov Chain Monte Carlo (MCMC) algorithm. Two major problems arose when using a conventional Metropolis-Hastings MCMC algorithm: the correlation between the normalized planetary mean radius $R_p/R_s$ and the $k_2$ (see Discussions), and the computing time. By recalling that one transit simulation takes about 450 ms, a random walk Metropolis of 3 chains, each consisting of 1 million steps, would take roughly 15 days without parallelizing the chains and 5 days with parallelization, assuming convergence would be reached. We found that a random walk Metropolis is not efficient in taking care of the previously mentioned correlation, leading to very long computing times. We solved this issue by using the so-called Differential Evolution Markov Chain (DE-MC) with snooker updater \citep{terbraak2008}. It solves the problem of choosing an appropriate scale and orientation for the jumping distribution by using the past states of each chains. \citet{cubillos2017} implemented a parallelized version of the algorithm in a Python library called MC$^3$. For each case we used 12 chains, each consisting of roughly 41500 steps (500,000 steps in total). The convergence was checked through a Gelman-Rubin test by ensuring that $\hat{R} < 1.01$ for every parameters \citep{gelman1992}. The peak of the posterior distribution defines the best solution, while the uncertainties are estimated by the $68\%$ confidence intervals.\\

\begin{deluxetable}{ccc}
\tablecaption{Adopted priors on the fitted parameters for the DE-MC. Three cases were considered for the limb darkening coefficients. \label{tab:priors}}
\tablehead{
\colhead{Parameter} & \colhead{Prior} & \colhead{Interval}
}
\startdata
$i$ & Uniform & [70.0; 90.0] \\
$u_p$ & Gaussian & $\mathcal{N}(0.6, [0.01, 0.005])$ \\
$u_m$ & Gaussian & $\mathcal{N}(0.0, [0.01, 0.005])$ \\
$d/R_s$ & Uniform & [3.0; 5.0] \\
$R_p/R_s$ & Uniform & [0.10; 0.20] \\
$k_2$ & Uniform & [0.0; 1.5]
\enddata
\end{deluxetable}

The adopted priors on the fitted parameters are reported in Table \ref{tab:priors}. In order to investigate the impact of stellar limb darkening uncertainty, we considered two cases: Gaussian priors with standard deviations of 0.01 and 0.005. In doing so, we can compare our results to recently published performances with ellipsoidal shape models \citep{akinsanmi2018}. A discussion about the accuracy of our present knowledge of limb darkening can be found in \citet{csizmadia2013}, which can be used as a baseline for the above choices of standard deviations.

\section{Discussion} \label{sec:discussion}

\begin{figure*}[ht!]
\plotone{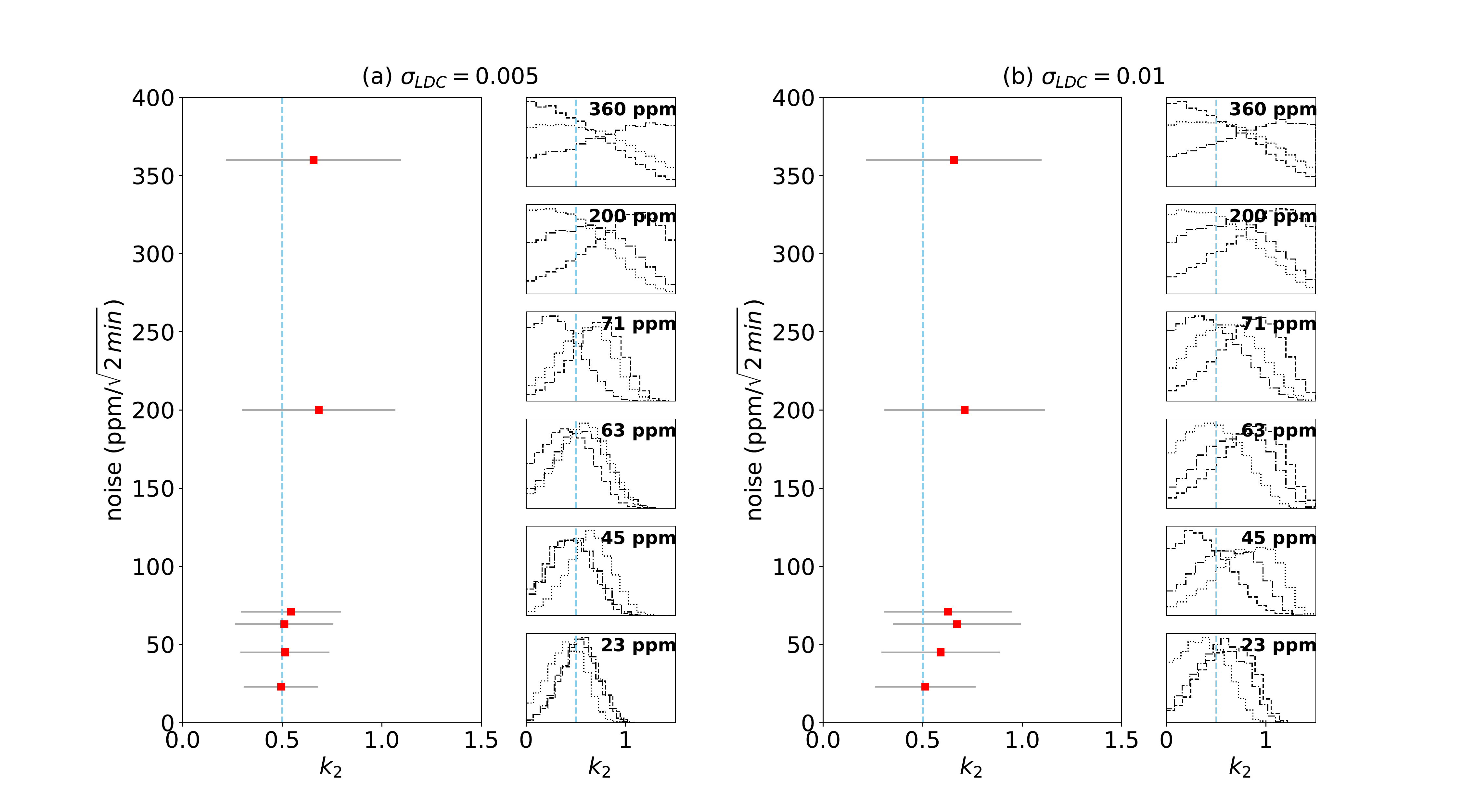}
\caption{(a) Left: average values of the three realizations as a function of the noise level. Right: $k_2$ posterior distributions. The considered standard deviation of the limb darkening coefficients was 0.005; (b) Same as (a) but for a standard deviation of the limb darkening coefficients equal to 0.01.}
\label{fig:results}
\end{figure*}

In Figure \ref{fig:results} we present the average values of the three realizations for each facility, in terms of mean and standard deviation of the measured $k_2$, for both considered priors on the limb darkening coefficients. We also show the posterior distributions of $k_2$ of all realizations to assess the quality of the parameter estimation. The measured value must be precise and accurate to confidently say that the model can retrieve $k_2$. Thus, we require a precision of at least $2\sigma$ and a relative error $|k_2-0.5|/0.5\leq5\%$.\\
\\
For a well constrained stellar limb darkening, we get a least a 2$\sigma$ detection with a relative error $< 5\%$, for noise levels up to 63 ppm/$\sqrt{2\,min}$. At 71 ppm/$\sqrt{2\,min}$ we also obtain a 2$\sigma$ detection of $k_2$, but with a relative error of about 9\%. For higher noise levels, the relative error drastically drops and the posterior distributions of $k_2$ widen and flatten, covering the whole physical range [0; 1.5] (see Figure \ref{fig:results}(a)).\\
When the accuracy on the limb darkening coefficients decreases (Figure \ref{fig:results}(b)), we are able to reliably recover $k_2$ with a noise level of 23 ppm/$\sqrt{2\,min}$ only. For higher noise values, the precision and relative error decidedly decrease. We present in Appendix \ref{sec:ld_effect} posterior distributions of all fitted parameters for few realizations at different noise levels, for both limb darkening priors.\\

We notice that the degeneracy between limb darkening (LD) and $k_2$ can disappear in the future due to progress in LD studies. Once LD is well known, limb darkening coefficients (LDCs) can be fixed or strict priors can be applied. The difference between observed and theoretically predicted LDCs were reported and summarized in, \textit{e.g.}, \citet{claret2008,claret2009}, and \citet{csizmadia2013}. \citet{csizmadia2013} also emphasized that stellar spots - surface manifestation of the stellar magnetic field - will modify the LDCs of theoretical tables, which has been confirmed by recent case studies \citep{maxted2018,maxted2019}. We also note that LDCs can be measured if the transit depth is at least 400 times larger than the photometric noise \citep{csizmadia2013}. Therefore, we think that the degeneracy between LD and $k_2$ is temporary, and will be removed by further LD studies with focus on the impact of stellar activity on LDCs.\\

\citet{akinsanmi2018} found, using a three-axis ellipsoidal shape model from \citet{correia2014}, that a noise level of 50 ppm/$\,\sqrt[]{1\,min}$ (or equivalently 36 ppm/$\,\sqrt[]{2\,min}$) or smaller is required to reliably recover $k_2$ (or equivalently $h_2$) from simulated transit light curves of WASP-103b and WASP-121b. For a V$_{mag}=10.4$ star, that would require around 40 transits and 27 transits with CHEOPS and PLATO, respectively (Table \ref{tab:noise}). We found in this paper that one can safely recover $k_2$ in transit light curves of WASP-121b with noise levels as high as about 65 ppm/$\,\sqrt[]{2\,min}$ (or equivalently 90 ppm/$\,\sqrt[]{1\,min}$), improving the performance by almost a factor 2. This would require only 13 observed transit with CHEOPS, and 10 observed transits with PLATO. To reach such performances, errors smaller than 0.01 on the limb darkening coefficients are required. This is in agreement with \citet{akinsanmi2018}.\\
\\
We highlight here that the maximum noise level at which one can safely retrieve $k_2$ depends on how strongly the planet deforms. Hence, for a hot Jupiter orbiting a little further from its Roche limit, \textit{e.g.} WASP-103b or WASP-12b, that maximum noise level should not be higher than 65 ppm/$\sqrt{2\,min}$.\\
\\
Our study solely considers the ideal case of pure white noise. In reality other noise sources may be present in the data, \textit{e.g.} stellar granulation \citep{chiavassa2017} or inhomogeneities in the stellar photosphere \citep{pont2007}, which would require lower noise levels (higher number of transits) to reach the precision in $k_2$ presented above. Such non-gaussian noises would be fitted through gaussian processes \citep{foreman2017} or wavelet analysis \citep{carter2009}.\\
\\
It is true that not all observing facilities are primarily designed to observe gas giants. However, based on the results presented above, one can propose to observe WASP-121b and similar targets with JWST, PLATO and CHEOPS. To perform such observations from ground, one would have to consider the absorption of light due to the Earth atmosphere, the scintillation, as well as the atmospheric variability (\textit{e.g.} turbulence, clouds, dust, pressure). These combined effects would inevitably increase the need for a larger telescope diameter and would require techniques such as adaptive optics to reach space-based noise levels.\\
\\
For outer planets however, not only the rotational rate is important but also the spin axis tilt. For instance, \citet{zhu2014} showed that the spin axis obliquity alone may have a big influence on the amplitude of oblateness-induced effects. Hence, a more exhaustive study of the parameter space is needed for outer planets where surface deformations are dominated by rotation. \\
\\
One may wish to apply the method presented in this paper to super Earths and Neptune-like planets. However, such objects being much smaller and less easily deformable (\textit{e.g.} layers of finite rigidity) than Jupiter-like planets, surface deformations are by consequences smaller, making the retrieval of $k_2$ harder.\\
\\
We assumed in our model a spherical star. In reality, the stellar disk is also non spherical due to the stellar rotation and tidal interactions with the planet (\textit{e.g.} \citet{barnes2009}). Such baseline variations of the stellar flux may have an impact on the planetary shape retrieval, and thus should be investigated.

\section{Conclusion}

Close-in planets are subjected to large tidal distortions, while outer planets remaining in fast rotational rate are subjected to large rotational distortions. These effects lead to surface deformations, for which we provide a complete three-dimensional analytical model. The departure from spherical shape induces distortions in transit light curves, whose magnitude depends on the planetary interior model, expressed through the fluid Love numbers $k_j$ (or equivalently $h_j$). Our shape model is valid for arbitrary orbital and rotational configuration or spin rates, and allows direct fitting of the \textit{true} planetary mean radius, and fluid Love number $k_2$. Considering the close-in hot Jupiter WASP-121b as a test case we showed that, under present understanding of stellar limb darkening, a precision $\leq$ 65 ppm/$\sqrt{2\,min}$ (equivalently 90 ppm/$\sqrt{1\,min}$) is required to reliably retrieve $k_2$. We thereby improve the performance by almost a factor 2 compared to results using three-axis ellipsoidal shape models. Such a noise level can be achieved by CHEOPS in 13 observed transits, by PLATO in 10 observed transits, and by JWST in only two observed transit. Kepler was able to reach that precision in 5 observed transits. However, TESS would reach such a precision after 320 observed transits. On the other hand, TESS may detect a new target around a brighter star than WASP-121, hence requiring a lower number of observed transits to reach the required precision for Love number measurements. A careful treatment of noise sources is critical to achieve reliable measurements of $k_2$, and any improvement on stellar limb darkening would increase the performances summarized above. Such measurements would allow to further constrain exoplanetary internal structures by comparing the measured $k_2$ to theoretical interior model expectations.\\

\acknowledgments
We acknowledge support from the DFG via the Research Unit FOR 2440 \textit{Matter under planetary interior conditions}.

\clearpage

\appendix

\section{Difference in shape models} \label{sec:shapemodels}
In Table \ref{tab:Corr} we compare some planetary shapes calculated by \citet{correia2014} with those obtained using our model, for which an initial value of $k_2$ is given as input: $k_{2,input}=0.5$ for gaseous planets and $k_{2,input}=1.0$ for rocky planets \citep{yoder1995}. Synchronous rotation is assumed for all cases ($F_p=1$). We define $R_{sub} = r(\theta=90^{\circ},\phi=0^{\circ})$, $R_{anti} = r(\theta=90^{\circ},\phi=180^{\circ})$, $R_{trail} = r(\theta=90^{\circ},\phi=90^{\circ})$, and $R_{pole} = r(\theta=0^{\circ},\phi=0^{\circ})$. The equivalence is $a \leftrightarrow R_{sub}$, $b \leftrightarrow R_{trail}$, and $c \leftrightarrow R_{pole}$.  We compute radii values up to three digits as this is a common precision in exoplanetary observations.

\startlongtable
\begin{deluxetable}{cc|ccc|cccc}
\tablecaption{Planetary shape comparison between \citet{correia2014} and our model. \label{tab:Corr}}
\tablehead{
\colhead{} & \colhead{} & \multicolumn{3}{c}{\citet{correia2014}} & \multicolumn{4}{c}{This work} \\
\colhead{Planets} & \colhead{$k_{2,input}$} & \colhead{$a$} & \colhead{$b$} & \colhead{$c$} & \colhead{$R_{sub}$} & \colhead{$R_{trail}$} & \colhead{$R_{pole}$} & \colhead{$R_{anti}$} \\
 & & \colhead{($R_{\oplus}$)} & \colhead{($R_{\oplus}$)} & \colhead{($R_{\oplus}$)} & \colhead{($R_{\oplus}$)} & \colhead{($R_{\oplus}$)} & \colhead{($R_{\oplus}$)} & \colhead{($R_{\oplus}$)} 
}

\startdata
WASP-19b & 0.5 & 17.3 & 15.4 & 14.8 & 17.108 & 15.120 & 14.489 & 17.038 \\
WASP-12b & 0.5 & 21.3 & 19.3 & 18.7 & 21.143 & 19.016 & 18.318 & 21.076 \\
WASP-103b & 0.5 & 18.5 & 17.0 & 16.5 & 18.456 & 16.781 & 16.231 & 18.403 \\
WASP-52b & 0.5 & 14.7 & 14.0 & 13.8 & 14.808 & 14.094 & 13.859 & 14.794 \\
CoRoT-1b & 0.5 & 17.2 & 16.5 & 16.2 & 17.352 & 16.543 & 16.276 & 17.332 \\
WASP-78b & 0.5 & 19.4 & 18.7 & 18.5 & 19.604 & 18.928 & 18.705 & 19.590 \\
WASP-48b & 0.5 & 19.0 & 18.4 & 18.2 & 19.266 & 18.592 & 18.369 & 19.253 \\
WASP-4b & 0.5 & 14.8 & 14.4 & 14.2 & 15.037 & 14.519 & 14.349 & 15.025 \\
HAT-P-23b & 0.5 & 15.5 & 15.1 & 14.9 & 15.698 & 15.254 & 15.107 & 15.687 \\
WASP-43b & 0.5 & 11.7 & 11.4 & 11.3 & 11.887 & 11.552 & 11.442 & 11.878 \\
WASP-18b & 0.5 & 17.0 & 16.8 & 16.7 & 13.133 & 13.056 & 13.031 & 13.131 \\
55 Cnc e & 1.0 & 2.23 & 2.18 & 2.16 & 2.062 & 2.022 & 2.009 & 2.062 \\
Kepler-78b & 1.0 & 1.32 & 1.21 & 1.18 & 1.277 & 1.178 & 1.145 & 1.276 \\
Kepler-10b & 1.0 & 1.49 & 1.47 & 1.46 & 1.486 & 1.466 & 1.459 & 1.486 \\
CoRoT-7b & 1.0 & 1.59 & 1.58 & 1.58 & 1.589 & 1.577 & 1.574 & 1.589
\enddata
\end{deluxetable}

\onecolumngrid
\section{Impact of stellar limb darkening} \label{sec:ld_effect}
We present posterior distribution, mean value and 68\% confidence interval of all fitted parameters for three different noise levels, for both considered limb darkening priors.

\begin{figure}[!ht]
\plottwo{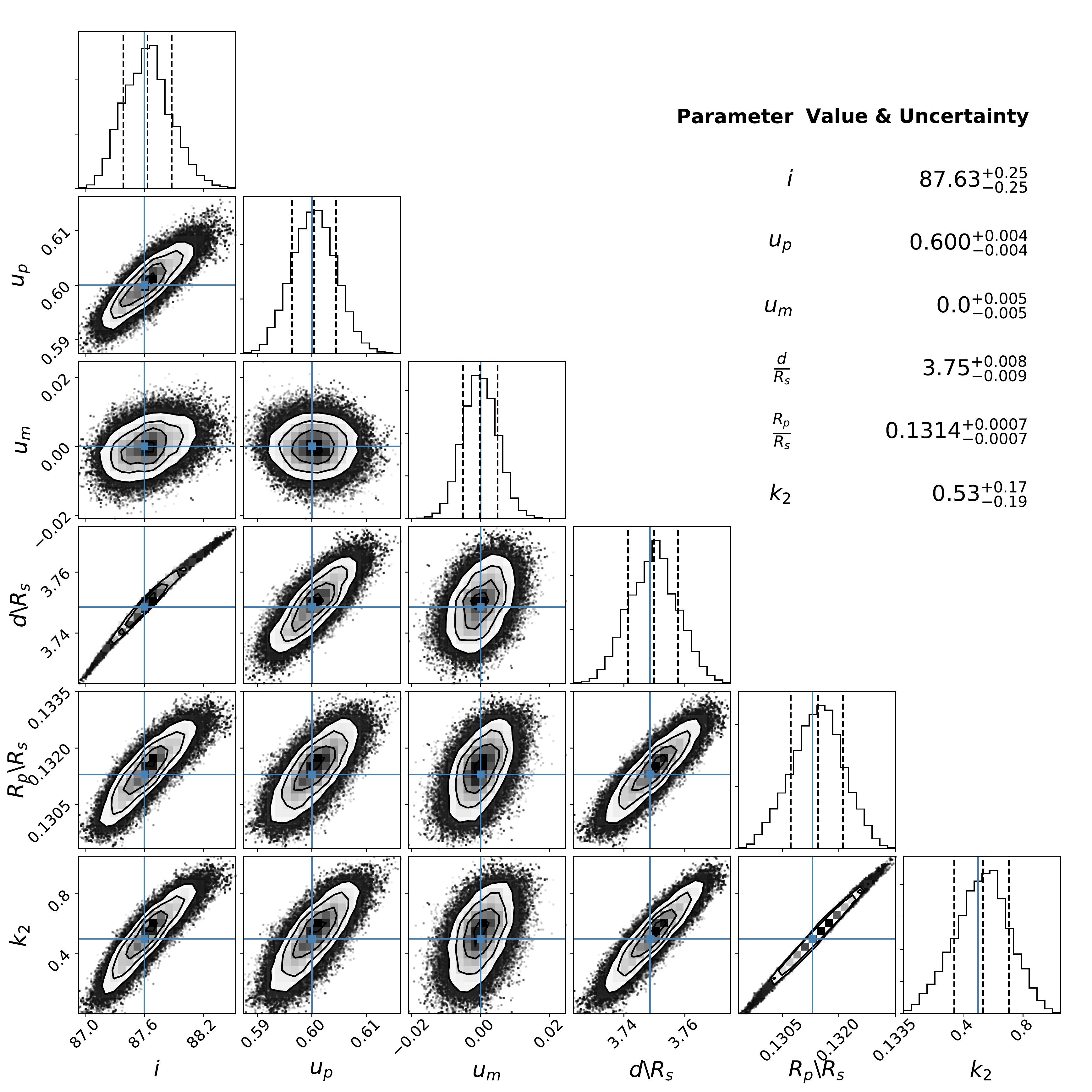}{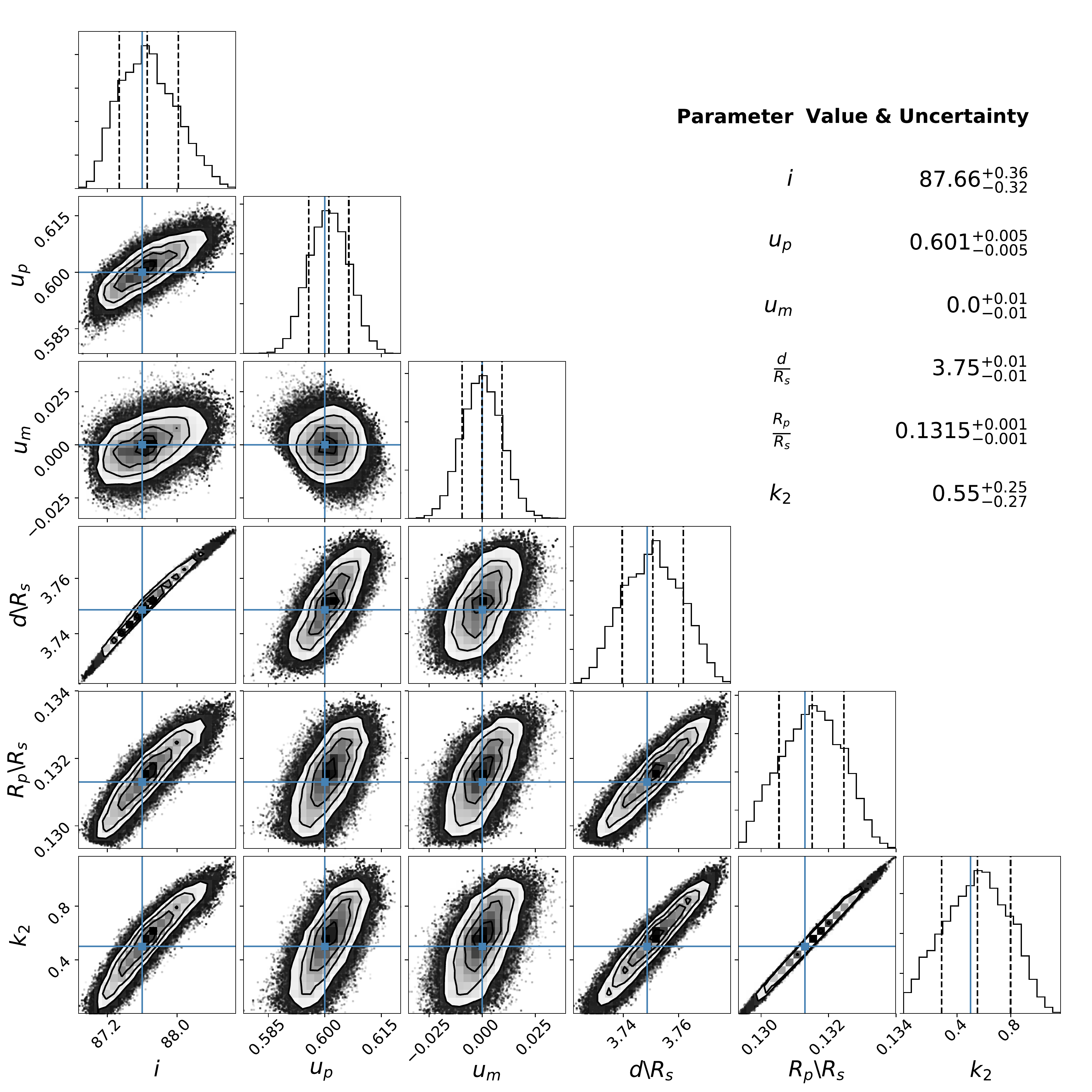}
\caption{Posterior distributions of the fitted parameters for a noise level of 23 ppm/$\sqrt{2\,min}$. Left: $\sigma_{LDC}=0.005$. Right: $\sigma_{LDC}=0.01$.}
\end{figure}

\begin{figure}
\plottwo{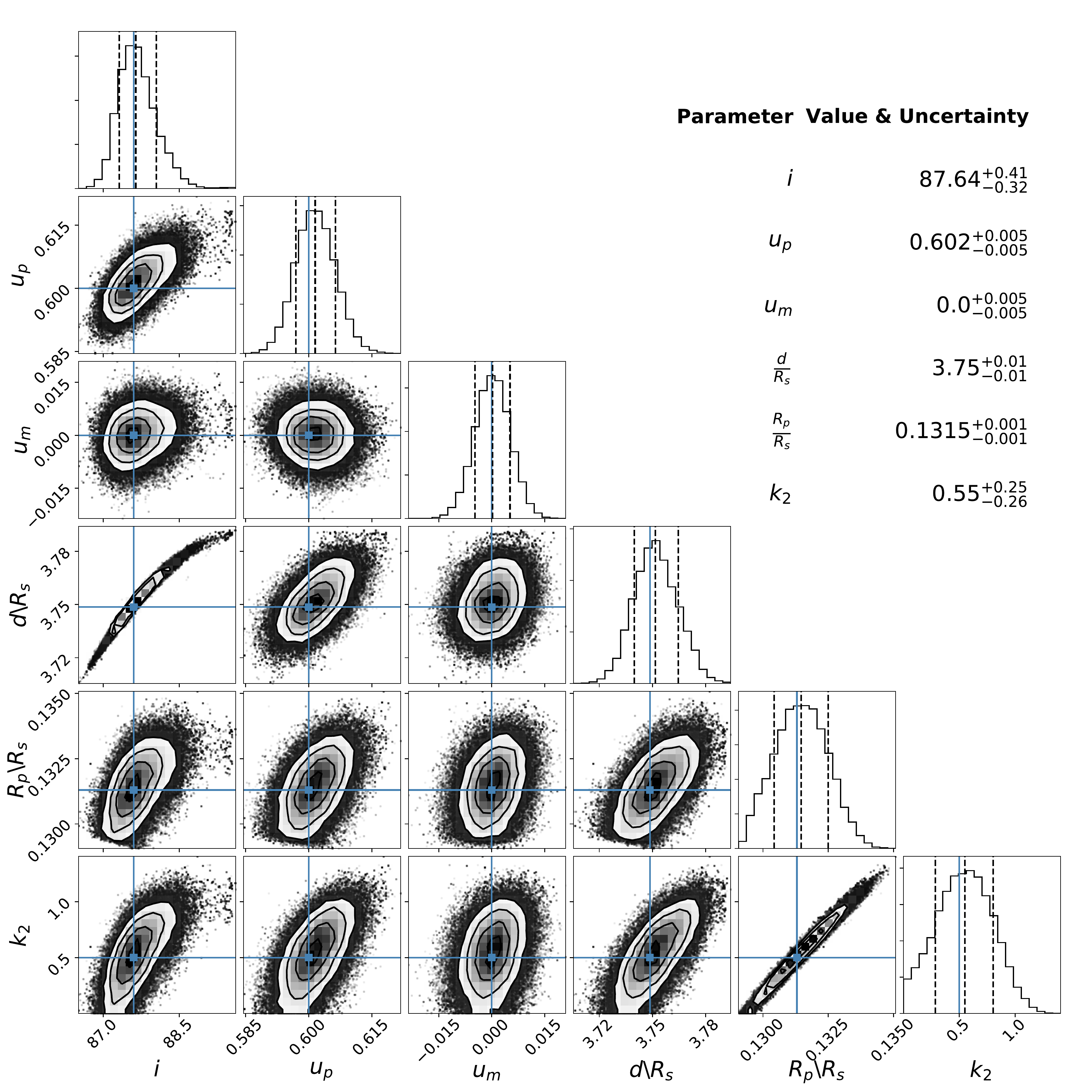}{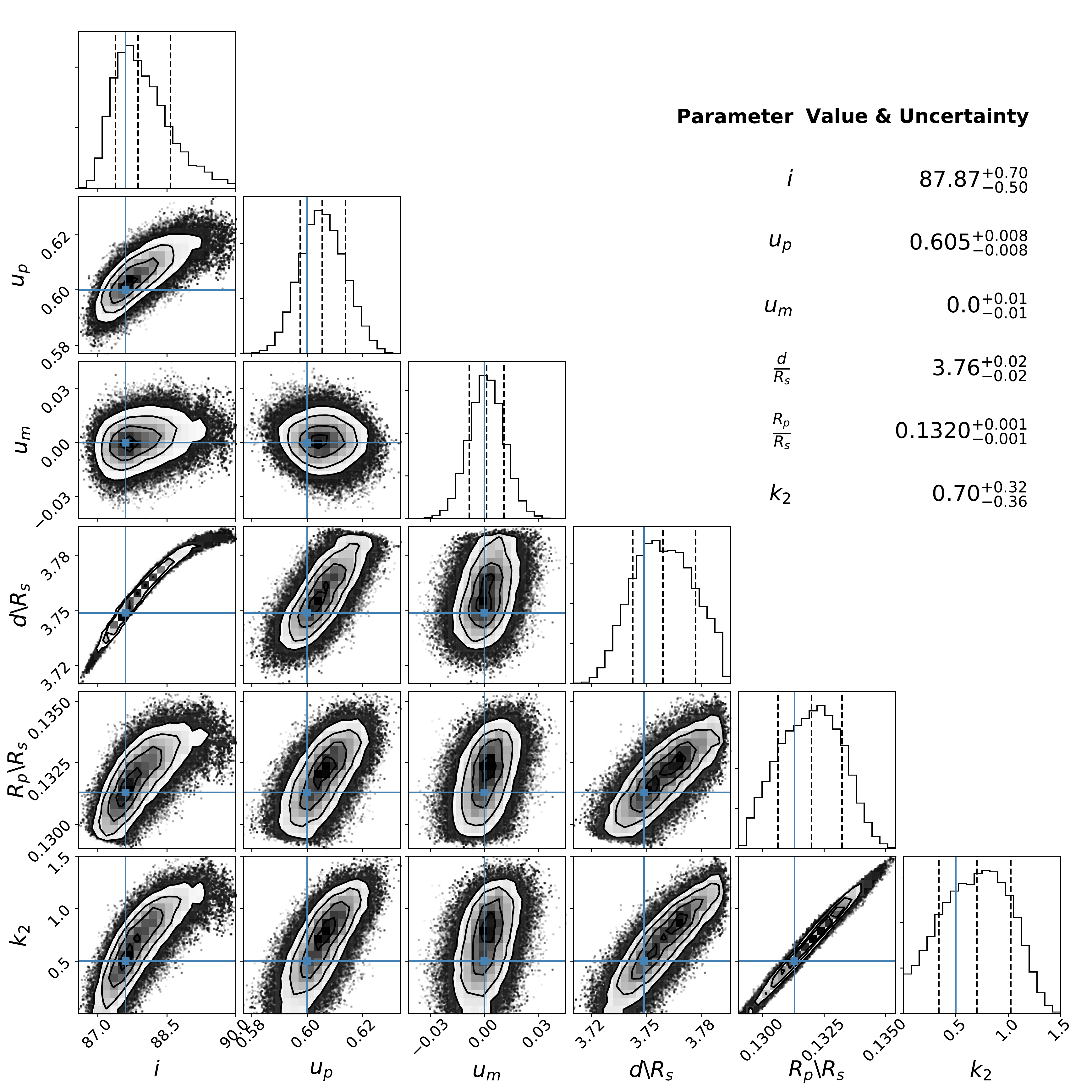}
\caption{Posterior distributions of the fitted parameters for a noise level of 63 ppm/$\sqrt{2\,min}$. Left: $\sigma_{LDC}=0.005$. Right: $\sigma_{LDC}=0.01$.}
\end{figure}

\begin{figure}
\plottwo{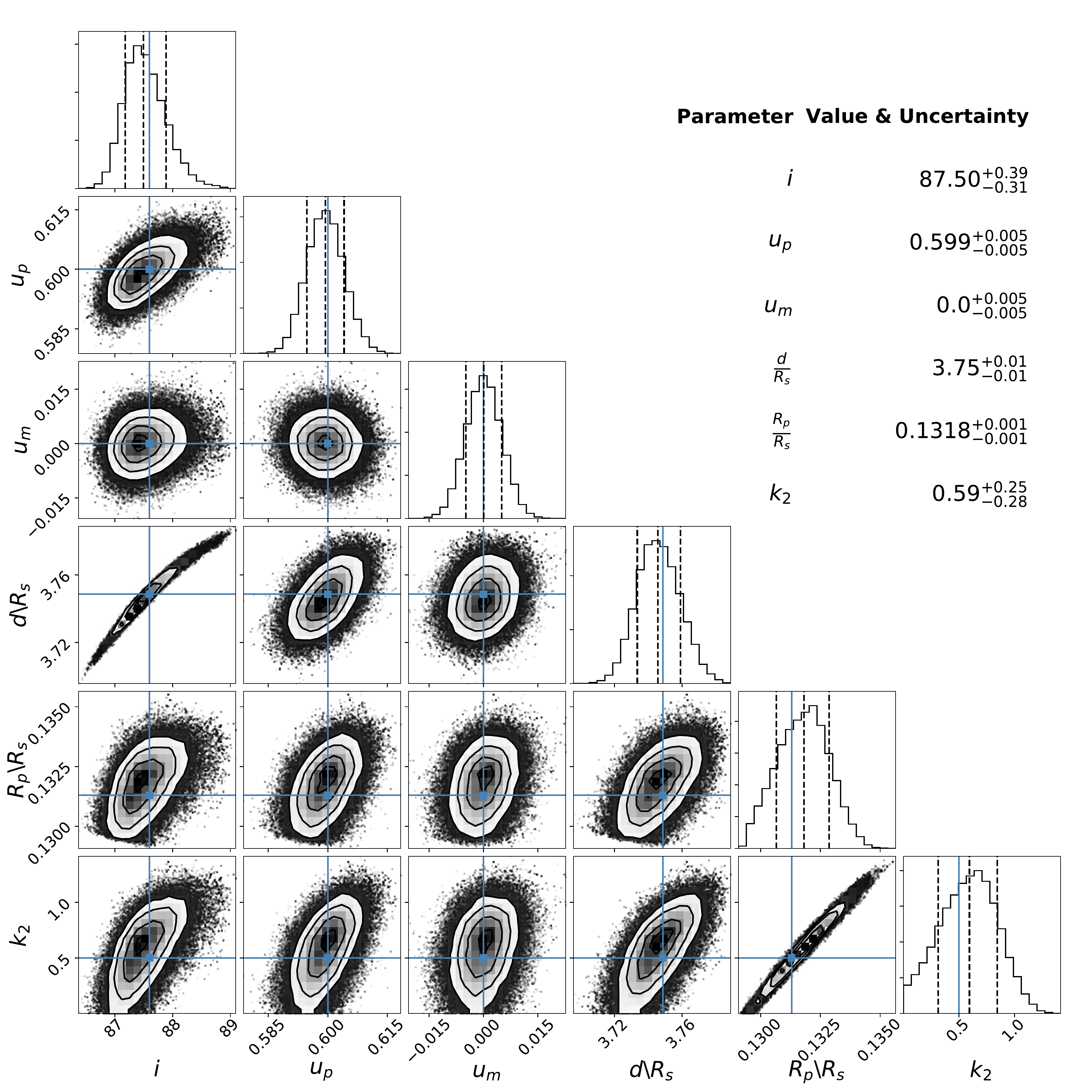}{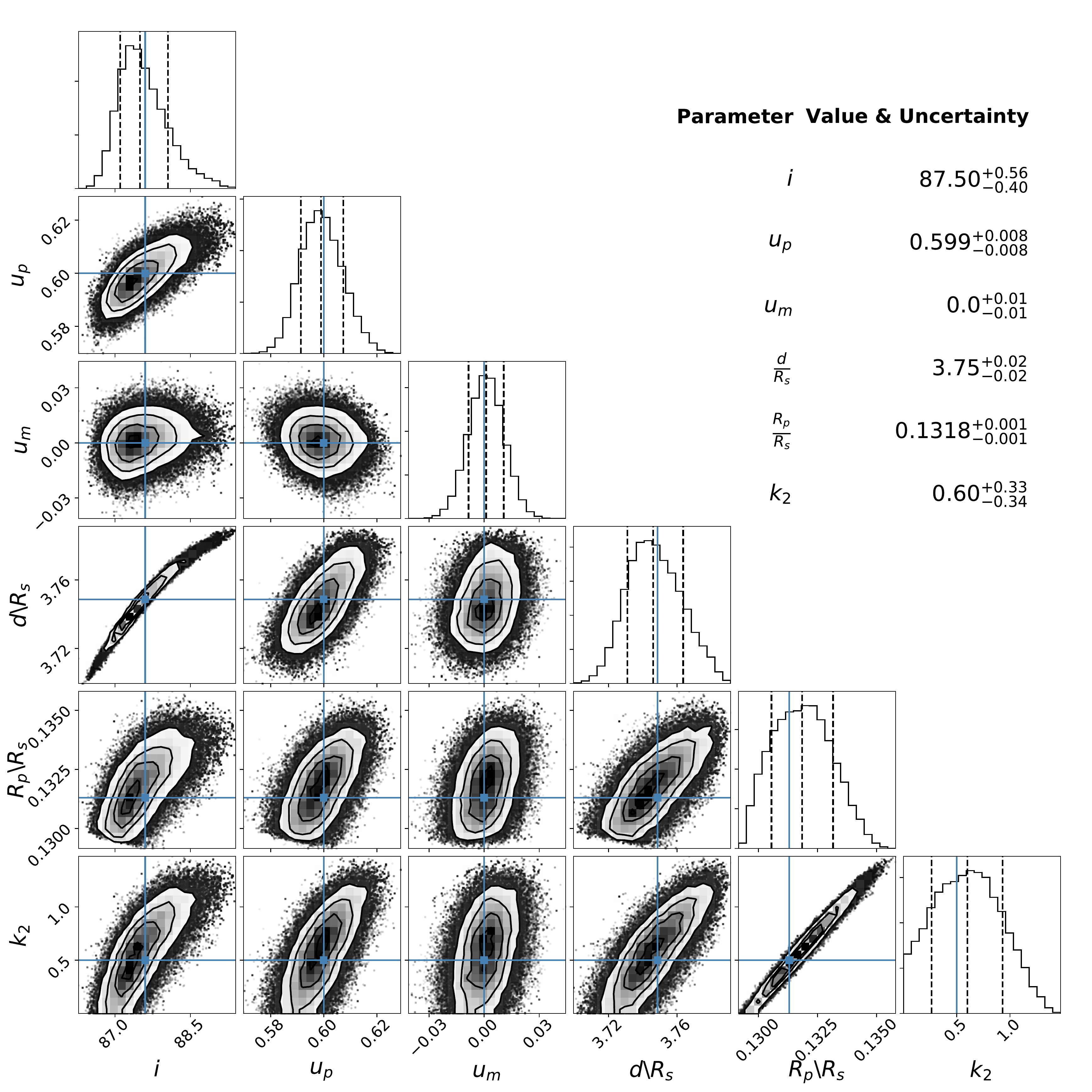}
\caption{Posterior distributions of the fitted parameters for a noise level of 71 ppm/$\sqrt{2\,min}$. Left: $\sigma_{LDC}=0.005$. Right: $\sigma_{LDC}=0.01$.}
\end{figure}

\end{document}